\documentclass[conference]{IEEEtran}
\IEEEoverridecommandlockouts

\usepackage[T1]{fontenc}        % proper hyphenation -> fewer justification gaps
\usepackage{cite}
\usepackage{amsmath,amssymb,amsfonts}
\usepackage{graphicx}
\usepackage{booktabs}
\usepackage{float}
\usepackage{textcomp}
\usepackage{xcolor}
\usepackage{url}
\usepackage[hidelinks]{hyperref}
\usepackage{tikz}               % code-drawn (editable) diagrams
\usetikzlibrary{arrows.meta}
\usepackage{microtype}          % glyph protrusion/expansion -> tight, even spacing
\newif\ifreview
\reviewfalse
\makeatletter
\newcommand{\linebreakand}{%
  \end{@IEEEauthorhalign}\hfill\mbox{}\par
  \mbox{}\hfill\begin{@IEEEauthorhalign}}
\makeatother

\begin{document}

%------------------------------------------------------------------- TITLE

\title{Clock-Gating Insertion Strategies on an Open-Source MSP430 Core:
A Reproducible PPA Study and a Gate-Level Simulation Caveat}

% \title{Clock-Gating Insertion Strategies on an Open-Source MSP430 Core:\\
% A Reproducible PPA Study and a Gate-Level Simulation Caveat}

% Double-blind: \reviewtrue -> anonymized authors; \reviewfalse -> real authors.
\ifreview
\author{\IEEEauthorblockN{Anonymous Author(s)}
\IEEEauthorblockA{Affiliations omitted for double-blind review}}
\else
\author{
\IEEEauthorblockN{Xingran Huang}
\IEEEauthorblockA{\textit{Dept.\ of Computer Science and Engineering} \\
\textit{University of California, Riverside}\\ Riverside, CA, USA\\
xhuan230@ucr.edu}
\and
\IEEEauthorblockN{Qiming Guo}
\IEEEauthorblockA{\textit{Dept.\ of Computing Sciences} \\
\textit{Texas A\&M University--Corpus Christi}\\ Corpus Christi, TX, USA\\
qguo2@islander.tamucc.edu}
\and
\IEEEauthorblockN{Jinwen Tang}
\IEEEauthorblockA{\textit{EECS Department} \\
\textit{University of Missouri}\\ Columbia, MO, USA\\
jt4cc@umsystem.edu}
\linebreakand
\IEEEauthorblockN{Wenqi Jia}
\IEEEauthorblockA{\textit{Dept.\ of Computer Science} \\
\textit{University of Texas at Arlington}\\ Arlington, TX, USA\\
wenqi.jia@uta.edu}
\and
\IEEEauthorblockN{Dongzheng Wang}
\IEEEauthorblockA{\textit{Dept.\ of Computer Science} \\
\textit{Boston University}\\ Boston, MA, USA\\
wdz1997@bu.edu}
}
\fi

%\maketitle

% ===== IEEE copyright notice (CAMERA-READY) =====================================
% ===== IEEE copyright notice (from UEMCON copyright info, Conf. Record #70256) =====
\IEEEpubid{\makebox[\columnwidth]{\footnotesize
  979-8-3195-2011-1/26/\$31.00~\copyright~2026 IEEE\hfill}%
  \hspace{\columnsep}\makebox[\columnwidth]{}}

\maketitle
%------------------------------------------------------------------- ABSTRACT
\begin{abstract}
Clock gating, the standard technique for cutting dynamic power, is introduced
either as hand-written behavioral clock gates at the register-transfer level
(RTL) or as integrated clock-gating (ICG) cells inserted automatically during
synthesis; the two are widely treated as interchangeable. In this paper we show, on a real open-source 16-bit
microcontroller core (openMSP430) synthesized with a 32\,nm standard-cell
library, that they are \emph{not} equivalent in practice: behavioral latch-based
RTL gating is functionally correct in ideal RTL simulation (10/10 self-checking
testcases, identical to the ungated baseline) yet \textbf{fails at gate level}:
the gated multiplier result is never captured and reads zero, while tool-inserted
ICG cells pass gate-level simulation cleanly (10/10). We root-cause the failure to
a hold race introduced by the late latch+AND gated clock, and show it is
\emph{consistently reproducible for the discrete latch+AND implementation under
the evaluated synthesis, timing, and simulation flow} (it persists across eight
simulation configurations including full Standard Delay Format (SDF)
back-annotation), not a simulator-setting artifact. We then quantify the
power/area/timing (PPA) impact of three gating strengths: RTL behavioral (Opt1),
synthesis ICG (Opt2), and both (Opt3), against the ungated baseline, across four
workloads and three process corners (ss/tt/ff). The benefit is corner-robust: ICG
(Opt2) cuts dynamic power by 74--81\% and total power by 25--30\% at every corner.
We also show that in this leakage-dominated 32\,nm regime the total-power win
comes from the area/leakage reduction that gating brings (leakage $-24$ to
$-30\%$), not from the large dynamic saving, which instead dominates active-mode
energy. Our recommendation for low-power design on open-source cores is to prefer
tool-inserted ICG cells over hand-written behavioral clock gates. The full flow
(Design Compiler synthesis, PrimeTime PX power, and self-checking verification)
is released as an open artifact.
\end{abstract}

\begin{IEEEkeywords}
openMSP430, clock gating, integrated clock gating, low-power design, dynamic
power, gate-level simulation, ASIC synthesis, PrimeTime.
\end{IEEEkeywords}

%==============================================================================
\section{Introduction}\label{sec:intro}
%==============================================================================
% If the IEEE copyright notice overlaps text at the bottom of page 1, uncomment:
\IEEEpubidadjcol
Dynamic power dominates the active-mode energy of digital designs, and clock
gating (suppressing the clock to registers that are idle) is the most widely
used technique to reduce it~\cite{yeap1998,wu2000}. In a
microcontroller such as the openMSP430, large blocks (the $16\times16$ hardware
multiplier, the watchdog, peripheral timers) are idle for much of normal
operation, making them natural clock-gating targets.

Clock gating is introduced into a design in two principal ways. (i) \emph{Behavioral}
RTL clock gates: a hand-written latch-plus-AND structure in the source that the
designer instantiates explicitly. (ii) \emph{Integrated clock-gating} (ICG) cells:
characterized standard cells that the synthesis tool inserts automatically (e.g.,
Design Compiler \texttt{compile\_ultra -gate\_clock})~\cite{emnett2000}. In
teaching material and in much practice these are treated as two equivalent routes
to the same result. We show they are not equivalent once the design is taken to
gate level on a real library.

Using the openMSP430~\cite{openmsp430} (an open-source, MSP430-compatible 16-bit
core) and a 32\,nm standard-cell library (SAED32, worst-case ss corner), we
implement and compare an ungated baseline and three gating strengths: RTL
behavioral gating (Opt1), synthesis-inserted ICG (Opt2), and both together
(Opt3). We verify each with 10 self-checking directed/corner testcases in both
ideal RTL and gate-level netlist simulation, and we measure post-synthesis
frequency, area, and PrimeTime~PX average power. Our central finding is a
reproducibility caveat: the behavioral RTL gate is \emph{functionally correct}
(ideal RTL passes 10/10, identical to baseline) but its netlist \emph{does not
simulate cleanly at gate level}: the gated result reads zero, whereas
tool-inserted ICG passes 10/10 at gate level.

\noindent\textbf{Contributions.} The contributions of this paper are:
\begin{enumerate}\itemsep2pt
  \item \textbf{A reproducible, real-library PPA study} of three clock-gating
        strengths on an open-source MCU core, reporting frequency, area,
        internal/dynamic power, and total power against an ungated baseline,
        across four workloads and three process corners (ss/tt/ff); we show the
        benefit is corner-robust and that, in this leakage-dominated regime, the
        total-power win comes from gating's area/leakage reduction
        (Section~\ref{sec:results}).
  \item \textbf{A gate-level simulatability caveat with root-cause analysis}:
        behavioral latch clock gates are RTL-correct but not
        gate-level-simulatable due to a gated-clock hold race; we show the result
        register captures zero despite 21{,}926 datapath gate-enable toggles (the
        CPU is not frozen), and that the failure persists across eight simulation
        configurations and full SDF back-annotation, while ICG cells (CGLPPRX2)
        are clean (Section~\ref{sec:caveat}).
  \item \textbf{A concrete design recommendation}: prefer tool-inserted ICG over
        hand-written behavioral clock gates for gate-level-clean low-power
        design, and an \textbf{open artifact} (RTL switch, full DC/PrimeTime flow,
        self-checking testbench) for reproduction (see the Artifact         Availability section).
\end{enumerate}

The rest of the paper is organized as follows.
Section~\ref{sec:related} positions this work within AI-driven applications
that motivate energy-efficient MCUs, and reviews related work on clock gating.
Section~\ref{sec:bg} gives background on the openMSP430 and the two gating

% [FIG 0] overview, drawn as editable TikZ code (edit coords / text / colors / styles below)
% coords in mm, y=0 is the mid-line; \resizebox auto-fits to column width so edits never overflow
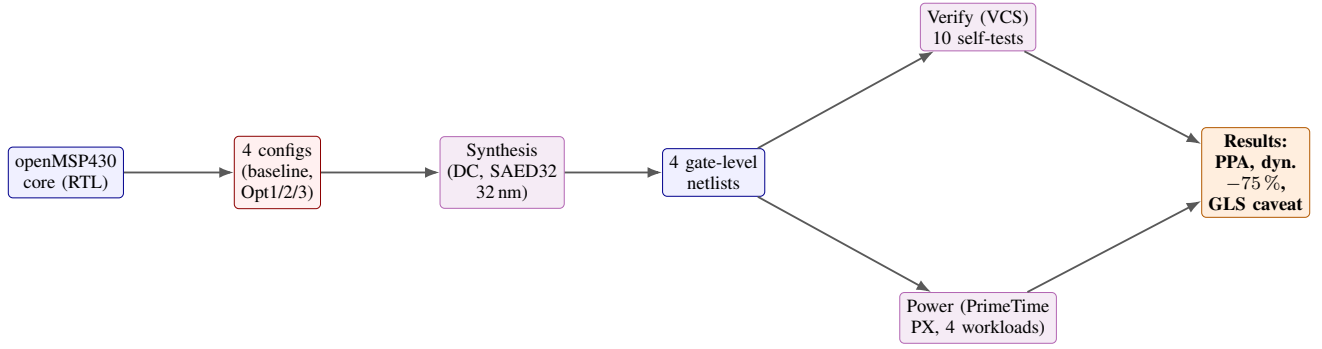
\begin{figure*}[t]
  \centering
  \resizebox{0.95\textwidth}{!}{%
  \begin{tikzpicture}[x=1mm,y=1mm,font=\footnotesize,
      core/.style={draw=blue!55!black,fill=blue!6, rounded corners=2pt,align=center,inner sep=3pt},
      cfg/.style ={draw=red!55!black, fill=red!6,  rounded corners=2pt,align=center,inner sep=3pt},
      tool/.style={draw=violet!60,   fill=violet!8,rounded corners=2pt,align=center,inner sep=3pt},
      res/.style ={draw=orange!70!black,fill=orange!13,rounded corners=2pt,align=center,inner sep=3pt,font=\footnotesize\bfseries},
      ar/.style  ={-{Latex[length=2.2mm]},thick,black!65}]
    % ---- nodes: \node[style] (name) at (x,y) {text}; edit position/text here ----
    \node[core] (c1) at (12,0)   {openMSP430\\core (RTL)};
    \node[cfg]  (c2) at (44,0)   {4 configs\\(baseline,\\Opt1/2/3)};
    \node[tool] (c3) at (78,0)   {Synthesis\\(DC, SAED32\\32\,nm)};
    \node[core] (c4) at (110,0)  {4 gate-level\\netlists};
    \node[tool] (v)  at (150,22) {Verify (VCS)\\10 self-tests};
    \node[tool] (p)  at (150,-22){Power (PrimeTime\\PX, 4 workloads)};
    \node[res]  (r)  at (192,0)  {Results:\\PPA, dyn.\\$-75\,\%$,\\GLS caveat};
    % ---- arrows ----
    \draw[ar] (c1) -- (c2);   \draw[ar] (c2) -- (c3);   \draw[ar] (c3) -- (c4);
    \draw[ar] (c4) -- (v);    \draw[ar] (c4) -- (p);
    \draw[ar] (v) -- (r);     \draw[ar] (p) -- (r);
  \end{tikzpicture}}
  \caption{Overview of the study. From the openMSP430 RTL we build four
  configurations (ungated baseline plus three clock-gating strengths), synthesize
  each with Design Compiler on a 32\,nm SAED32 library into a gate-level netlist,
  and evaluate every netlist two ways: functional verification in VCS gate-level
  simulation (GLS) and average-power analysis in PrimeTime~PX across four
  workloads. The outcome is a PPA comparison, a 75\% dynamic-power reduction, and
  a gate-level simulatability caveat.}
  \label{fig:overview}
\end{figure*}

%==============================================================================

%==============================================================================
%==============================================================================%==============================================================================
\section{Related Work}\label{sec:related}
%==============================================================================
\subsection{AI-driven applications across domains}
Propelled by deep learning~\cite{lecun2015deep} and Transformer-based large
language models~\cite{vaswani2017attention,brown2020gpt3}, AI has expanded
rapidly into engineering practice. Early applications targeted building energy
management: occupant-behavior-integrated air-conditioning
simulation~\cite{xie2019distributedac}, regional building-load
prediction~\cite{zhou2021regionalbuildingload}, deep-learning thermal-load
meta-modeling~\cite{zhou2022deepthermal}, and KNN surrogates for long-term
thermal-load prediction~\cite{liang2023surrogate}. A parallel systems-level line
makes AI and scientific computing themselves efficient: learning-enhanced lossy
compression of scientific
data~\cite{jia2024gwlz,jia2024neurlzarxiv,jia2025neurlz}, cross-field prediction
for compression~\cite{liu2024crossfieldscw,liu2025crossfieldhpdc}, dedicated
compression hardware~\cite{jia2025flare}, and model compression and acceleration
spanning mixture-of-experts pruning~\cite{yang2024moei2}, structured sparse
training~\cite{xiao2024transpa}, real-time vision transformers on mobile
devices~\cite{shu2024ecpvit}, Gaussian-splatting simplification and distributed
training~\cite{zhang2025gaussianspa,jia2026splaxel}, and ultra-light
vision-language adaptation~\cite{huang2025adaring}.

\subsection{AI for infrastructure, end users, and trustworthiness}
AI has since been deployed on cyber-physical infrastructure itself:
spatio-temporal graph neural networks model hydraulic dependencies in urban
wastewater networks~\cite{guo2024hydronet} and enable sparse-sensor forecasting
and monitoring of urban water systems~\cite{guo2025watermonitor}, while
AquaSentinel couples an MoE ensemble with LLM agents for pipeline anomaly
detection~\cite{guo2025aquasentinel}. In parallel, LLM-driven human-centered
systems bring AI to end users, including adaptive mental-health support and
assessment~\cite{guo2024soullmate,guo2024soullmateapp}, custom-GPT psychological
pre-screening~\cite{tang2024prescreening}, adaptive campus-climate survey
chatbots~\cite{tang2025tigergpt}, and integrated campus well-being
tools~\cite{tang2026wellbeing}. Growing deployment raises trustworthiness
concerns, motivating machine unlearning for spatio-temporal
graphs~\cite{guo2025stgunlearning,guo2025unlearningsa} and studies of attacks on
LLMs and AI agents~\cite{guo2025aea}.

\subsection{The demand for energy-efficient embedded computing}
These applications increasingly reach embedded and IoT endpoints such as sensor
nodes and on-device inference platforms. Edge intelligence pushes AI from the
cloud toward such endpoints~\cite{zhou2019edge}, MCUNet runs deep-learning
inference directly on microcontrollers~\cite{lin2020mcunet}, and MLPerf Tiny
standardizes ML benchmarking on ultra-low-power
hardware~\cite{banbury2021mlperftiny}. There the energy budget of MCU-class
cores is decisive: near-threshold operation is combined with fine-grained gating
to cut active-mode energy~\cite{gautschi2017}, and clock gating is the
lowest-cost circuit-level lever, which we review next.

\subsection{Clock-gating fundamentals and low-power design}
Clock gating is a long-established dynamic-power technique covered in low-power
design texts~\cite{yeap1998,pedram1996power} and power-management
surveys~\cite{benini2000survey}; recent surveys catalog the full design space
from simple gate- and latch-based gating to data-driven and look-ahead
schemes~\cite{bharathi2025survey}, with comparative studies quantifying their
power and area trade-offs~\cite{sahu2020comparative}. Early automatic approaches
synthesized gated clocks for finite-state machines by detecting highly probable
idle conditions~\cite{benini1996}, and Wu \emph{et al.} formalized clock gating
of sequential circuits with a quaternary clock model, deriving a per-flip-flop
gated clock that is synchronous with the master clock~\cite{wu2000}. More recent
work pushes activity-aware gating further: look-ahead clock gating computes each
flip-flop's enable one cycle ahead to relax timing while widening gating
coverage~\cite{wimer2014}, and circuit-level gated flip-flops cut clocking power
at the cell level~\cite{strollo2000}.

\subsection{Clock-gate placement and clock-tree power}
Beyond register-level gating, the placement and merging of clock-gating cells
affect clock-tree power; Teng and Soin optimize clock-gate locations during
clock-tree synthesis~\cite{teng2010}. This line of work targets the physical
clock network and is complementary to our register-level RTL-vs-ICG comparison.

\subsection{Clock gating on processors and microcontrollers}
Clock gating has been applied to complete embedded controllers: Kamaraju
\emph{et al.} report a 33\% total-power reduction on an FPGA-implemented
programmable RISC controller via clock gating~\cite{kamaraju2010}. We study the
same technique on the open-source openMSP430 core~\cite{openmsp430}, a widely
used MSP430-compatible soft core, and add a gate-level simulatability analysis
absent from prior applied studies.

\subsection{RTL vs.\ synthesis-inserted gating}
Two insertion routes are standard. Emnett and Biegel describe automatic RTL
clock-gating insertion in a commercial synthesis flow (Synopsys Power Compiler),
reporting up to a two-thirds power reduction on a 200k-gate ASIC while preserving
scan testability~\cite{emnett2000}; more recent work measures clock-gating
efficiency and its power impact directly within the synthesis
flow~\cite{attaoui2021synthesis}. Our Opt2 uses this tool-inserted ICG route,
whereas Opt1 uses a hand-written behavioral latch gate. Prior work compares such
strategies for power and area, but \emph{does not report a gate-level
simulatability difference between behavioral latch gates and tool-inserted ICG
cells on a real standard-cell library}: the gap this paper fills.

\section{Background}\label{sec:bg}
%==============================================================================
\subsection{The openMSP430 core}
The openMSP430 is a compact (\textasciitilde8{,}000-gate) 16-bit microcontroller
core written in synthesizable Verilog and binary-compatible with the TI MSP430
architecture~\cite{openmsp430}. It integrates an execution unit (ALU and register
file R0--R15), a memory-mapped $16\times16$ hardware multiplier, a memory
backbone, a basic clock module (MCLK/SMCLK/ACLK and low-power modes), a watchdog,
SFRs, and a serial debug interface. The synthesized device-under-test is the
top-level \texttt{openMSP430} module. A generic clock-gate cell
(\texttt{omsp\_clock\_gate.v}, a latch+AND) is available in the source.

% [FIG 1] block diagram + clock-gating insertion points
\begin{figure}[t]
  \centering
  \includegraphics[width=\columnwidth]{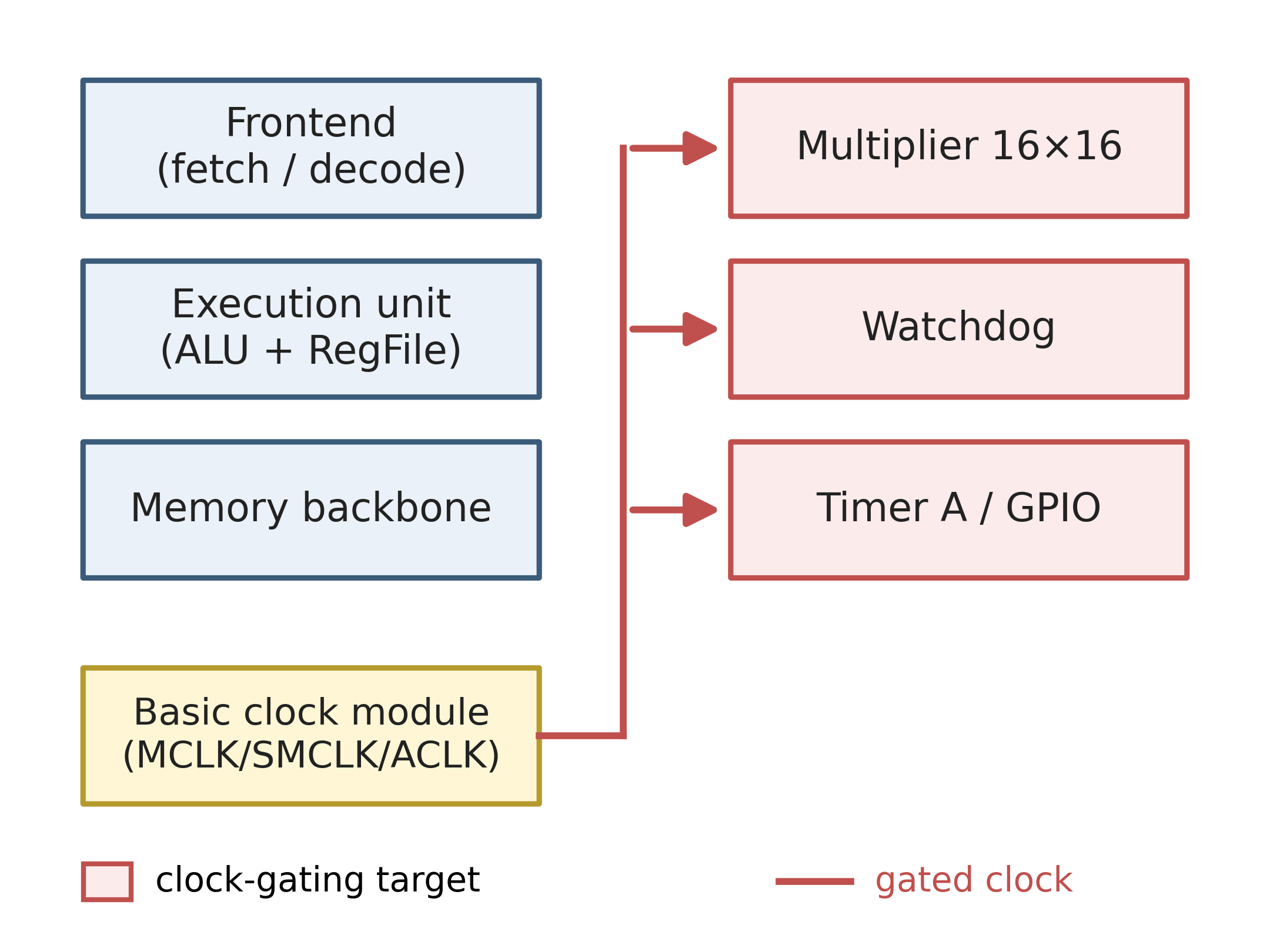}
  \caption{openMSP430 (simplified). Shaded blocks (multiplier, watchdog,
  timer/GPIO) are idle much of the time and are the clock-gating targets; red
  paths are gated clocks from the basic clock module.}
  \label{fig:block}
\end{figure}

\subsection{Two ways to insert clock gating}
% [FIG 2] the two clock-gate structures
\begin{figure}[t]
  \centering
  \includegraphics[width=\columnwidth]{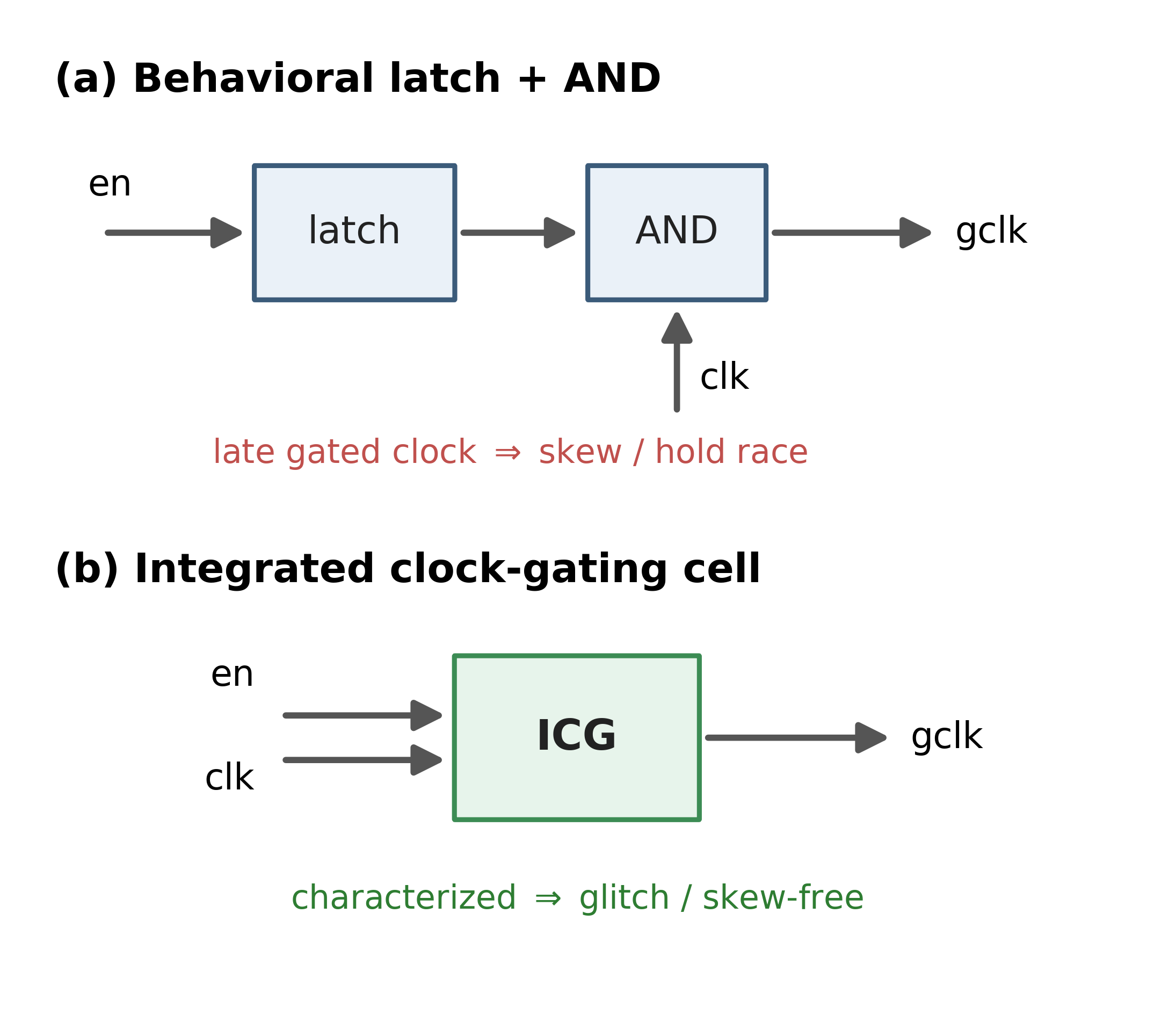}
  \caption{Behavioral latch+AND clock gate (a) vs.\ an integrated clock-gating
  (ICG) cell (b). The hand-written latch+AND produces a \emph{late} gated clock
  (race source); the ICG cell is characterized to avoid it.}
  \label{fig:gates}
\end{figure}
\emph{Behavioral} gating uses a hand-written latch+AND: the enable is latched on
the inactive clock edge and ANDed with the clock to produce a gated clock. The
\emph{ICG} route lets the synthesis tool replace this with a characterized
integrated clock-gating standard cell that is designed to avoid glitches and skew.
These are usually treated as equivalent; Section~\ref{sec:caveat} shows they are
not at gate level.

%==============================================================================
\section{Methodology}\label{sec:method}
%==============================================================================
Fig.~\ref{fig:overview} summarizes the end-to-end flow: four configurations,
synthesis into four netlists, and a two-pronged evaluation (gate-level
verification and PrimeTime power).

\subsection{Baseline and three gating implementations}
The upstream RTL only enables fine-grained gating inside an \texttt{`ifdef ASIC}
block, and \texttt{`ASIC} is undefined, so the baseline contains no active clock
gating. We define the baseline explicitly as this \emph{ungated} configuration so
the before/after comparison is fair; no functional edits are made to the baseline
RTL. We then build three variants:
\begin{itemize}\itemsep2pt
  \item \textbf{Opt1 -- RTL gating}: enable the source behavioral latch+AND gate
        (\texttt{omsp\_clock\_gate.v}) throughout the core, via a single
        \texttt{CG\_ENABLE} switch.
  \item \textbf{Opt2 -- synthesis gating}: Design Compiler inserts library ICG
        cells via \texttt{compile\_ultra -gate\_clock} with
        \texttt{set\_clock\_gating\_style}; no RTL change.
  \item \textbf{Opt3 -- both}: behavioral RTL gating and synthesis ICG together.
\end{itemize}
Because each module includes \texttt{openMSP430\_undefines.v} (which clears
command-line defines), defining \texttt{CG\_ENABLE} re-asserts \texttt{`CLOCK\_GATING}
on every per-module include; the optimized configuration is selected by a renamed
wrapper \texttt{openMSP430\_defines\_gated.v} so switching one included file flips
the whole design between baseline and optimized.

\subsection{Synthesis flow}
Synthesis uses Design Compiler with worst-case corner libraries
(\texttt{saed32rvt/lvt/hvt\_ss0p75v125c.db}, mixed RVT/LVT/HVT). A single
parameterized script synthesizes the baseline and each optimization separately,
producing four netlists. Maximum frequency is found by a clock-period sweep: the
period is reduced until worst slack would go negative, and the smallest period
that still closes timing (non-negative slack) is taken as $F_{max}$.

\subsection{Power measurement}
Average power is measured in PrimeTime~PX (averaged mode). For each netlist, a
workload is run in gate-level simulation to produce a value change dump (VCD);
PrimeTime annotates this switching activity onto the netlist and reports average
power. To avoid drawing conclusions from a single hand-picked stimulus, we
evaluate \emph{four} workloads with distinct activity profiles, generated as
MSP430 program images (\texttt{gen\_multi\_workloads.py}): \textbf{mult}
(sustained $16\times16$ multiplies, multiplier and datapath busy), \textbf{idle}
(watchdog held, no multiplies), \textbf{wdt} (free-running watchdog, exercising
the watchdog block), and \textbf{mix} (multiply bursts separated by idle gaps,
i.e.\ duty-cycled). The same four workloads are applied identically to all four
configurations. To test corner sensitivity, we additionally re-characterize power
at three corners by re-linking the corner libraries: \texttt{ss} (0.75\,V,
125\textcelsius), \texttt{tt} (1.05\,V, 25\textcelsius), \texttt{ff} (1.16\,V,
25\textcelsius), on the same netlists, since switching activity is
corner-independent.

\subsection{Self-checking verification}
The testbench instantiates the top \texttt{openMSP430} DUT with behavioral
program/data memories. Each testcase is a short MSP430 program that stores the
final hardware-multiplier result (\texttt{RESLO}/\texttt{RESHI}) into data memory;
the testbench reads it back and compares against a Python golden model of the
multiplier. Because the result is observed through the data-memory interface
(not by hierarchical probing), the same flow works for ideal RTL and flattened
netlist simulation. We wrote 10 directed/corner testcases targeting the gated
blocks, including idle$\rightarrow$active transitions, maximum-toggle operands,
signed multiplies, and carry into the high word.

%==============================================================================
\section{Results}\label{sec:results}
%==============================================================================
\subsection{Functional verification}
In \emph{ideal} RTL simulation all 10 testcases pass on the baseline and on all
three gated designs, with identical golden results, confirming that clock gating
changes power, not function. In \emph{netlist} simulation the baseline and Opt2
(synthesis ICG) pass all 10; Opt1/Opt3 do not (Section~\ref{sec:caveat}).

\begin{table}[H]
\caption{Functional pass rates (10 self-checking testcases, real VCS runs).}
\label{tab:pass}\centering\footnotesize
\begin{tabular}{@{}lcc@{}}
\toprule
\textbf{Configuration} & \textbf{Ideal RTL} & \textbf{Netlist} \\
\midrule
Baseline (ungated)          & 10/10 & 10/10 \\
Opt1 RTL gating             & 10/10 & fails (Sec.~\ref{sec:caveat}) \\
Opt2 synthesis gating (ICG) & 10/10 & 10/10 \\
Opt3 both                   & 10/10 & fails (Sec.~\ref{sec:caveat}) \\
\bottomrule
\end{tabular}
\end{table}

\subsection{PPA comparison}\label{sec:ppa}
% [REAL] data from final/evidence/part3_ppa_comparison.txt
Table~\ref{tab:ppa} compares frequency, area, and power (real DC / PrimeTime
data). All three gating variants reduce total power versus the baseline:
$-22.2\%$ (Opt1), $-29.2\%$ (Opt2), and $-23.1\%$ (Opt3). The dynamic power, which
clock gating targets, drops the most, from 49.6\,\textmu W (baseline) to
12.6\,\textmu W (Opt2). Clock gating also reduces area (shared gates replace
per-register feedback multiplexers), which lowers leakage at this corner; the cost
is a small $F_{max}$ reduction from the added gate in the clock path.
(Opt1/Opt3 fail the gate-level \emph{functional} check, but their power figures
here and in Table~\ref{tab:mwl} remain valid: the failure is confined to the
result-register capture, not the switching activity that sets average power;
see Section~\ref{sec:caveat}.)

\begin{table}[H]
\caption{PPA comparison (real Design Compiler / PrimeTime PX data).}
\label{tab:ppa}\centering\footnotesize
\begin{tabular}{@{}lccccc@{}}
\toprule
\textbf{Config} & \textbf{Freq.} & \textbf{Area} & \textbf{Power} & \textbf{Dyn.} & \textbf{vs} \\
 & (MHz) & (\textmu m\textsuperscript{2}) & (mW) & (\textmu W) & \textbf{base} \\
\midrule
Baseline       & 357.1 & 19474.0 & 1.484 & 49.64 & ref \\
Opt1 RTL       & 333.3 & 18043.2 & 1.155 & 39.30 & $-22.2\%$ \\
Opt2 synth ICG & 294.1 & 17745.1 & 1.050 & 12.58 & $-29.2\%$ \\
Opt3 both      & 333.3 & 18468.6 & 1.141 & 31.22 & $-23.1\%$ \\
\bottomrule
\end{tabular}
\end{table}

% [FIG 3] PPA bar charts (real data)
\begin{figure}[t]
  \centering
  \includegraphics[width=\columnwidth]{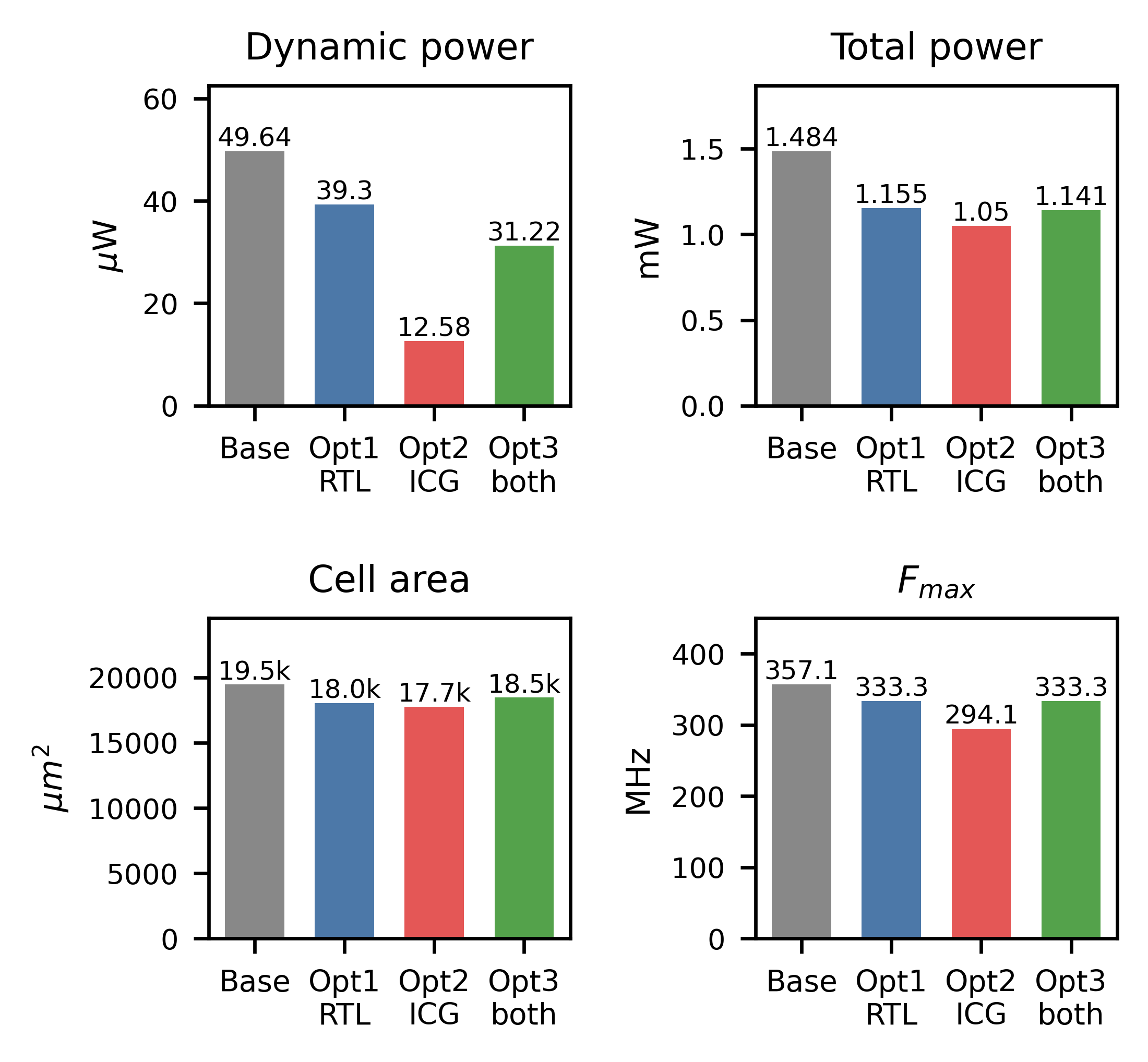}
  \caption{PPA across configurations (real DC/PrimeTime data): dynamic power,
  total power, cell area, and $F_{max}$. Dynamic power is the gating target and
  drops most (49.6\,$\rightarrow$\,12.6\,\textmu W for Opt2).}
  \label{fig:ppa}
\end{figure}

\subsection{Multi-workload power}\label{sec:multiwl}
% [REAL] data from final/evidence/part3_multiworkload_power.txt
Table~\ref{tab:mwl} reports dynamic power for all four configurations under the
four workloads. Two effects stand out. First, the dynamic-power reduction is
\emph{consistent across workloads}, not an artifact of one stimulus: ICG (Opt2)
cuts dynamic power by roughly 75\% for every workload. Second, the ungated
baseline is essentially workload-insensitive ($\approx$49.6\,\textmu W for mult,
idle, and mix) because every register is clocked each cycle regardless of
activity; only the free-running-watchdog workload (wdt) raises it (to
60.8\,\textmu W) by adding real toggling. The wdt workload, which exercises the
watchdog (a gated block), is the highest-power case for every configuration, and
gating still cuts its dynamic power substantially (60.8\,$\rightarrow$\,23.8\,\textmu W
for Opt2), confirming the gated block is active and benefits.

\begin{table}[H]
\caption{Dynamic power (\textmu W) per configuration and workload (PrimeTime PX, real).}
\label{tab:mwl}\centering\footnotesize
\begin{tabular}{@{}lcccc@{}}
\toprule
\textbf{Config} & \textbf{mult} & \textbf{idle} & \textbf{wdt} & \textbf{mix} \\
\midrule
Baseline       & 49.64 & 49.64 & 60.79 & 49.64 \\
Opt1 RTL       & 23.01 & 24.01 & 35.01 & 23.01 \\
Opt2 synth ICG & 12.58 & 13.51 & 23.78 & 12.58 \\
Opt3 both      & 31.22 & 30.94 & 45.19 & 31.22 \\
\bottomrule
\end{tabular}
\end{table}

\begin{figure}[t]
  \centering
  \includegraphics[width=\columnwidth]{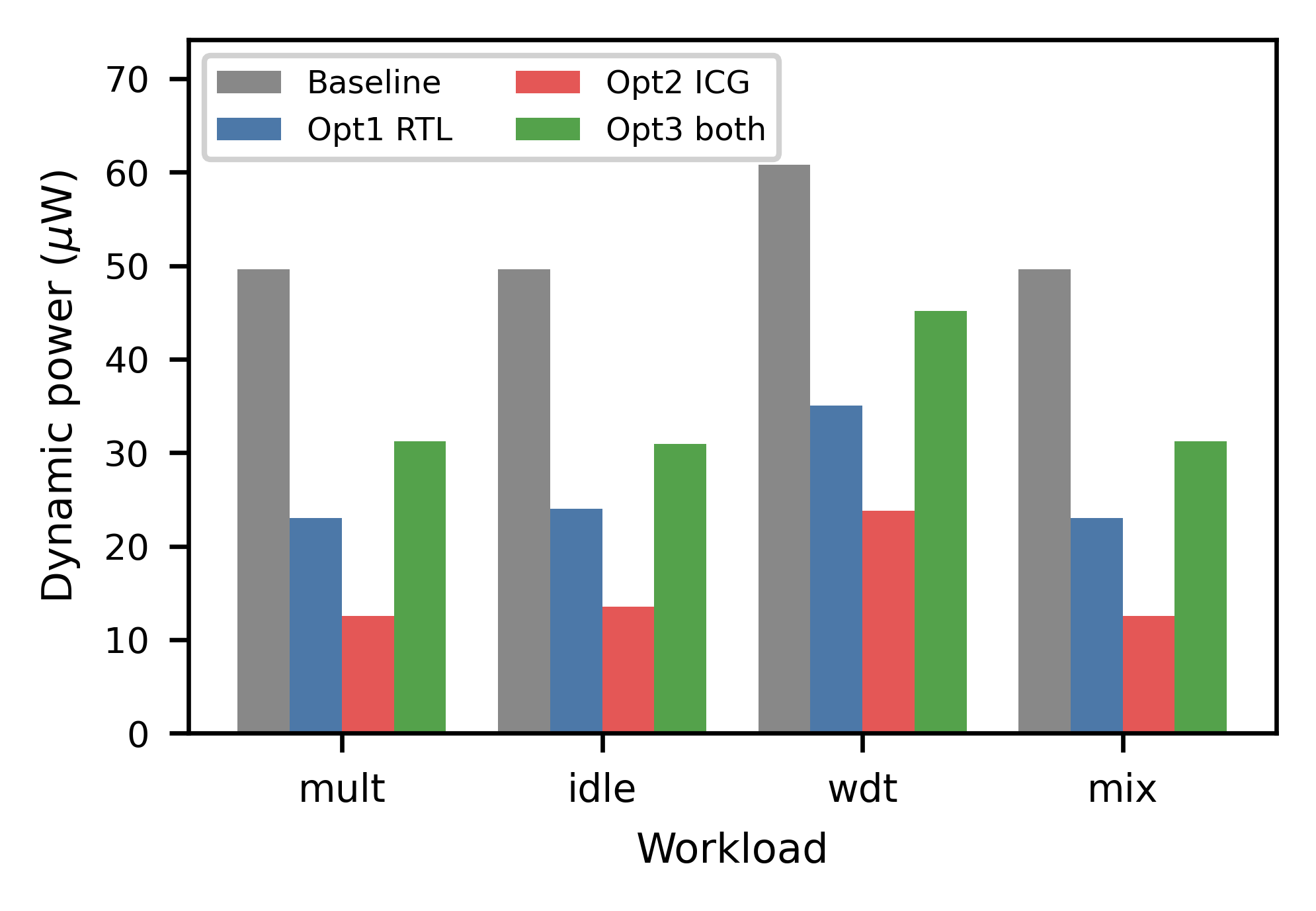}
  \caption{Dynamic power across four workloads (real PrimeTime PX). The gating
  saving is consistent across workloads; the free-running-watchdog workload (wdt)
  is the highest-power case for every configuration.}
  \label{fig:multiwl}
\end{figure}

\subsection{Where the saving comes from: clock-network power}\label{sec:perblock}
% [REAL] data from reports/power_group_<cfg>_w_mult.rpt
To attribute the saving, Table~\ref{tab:grp} breaks out the
\emph{clock-network} internal power (PrimeTime power groups, mult workload). For
the ungated baseline the clock-network internal power (49.63\,\textmu W) accounts
for essentially the entire dynamic power (49.64\,\textmu W): in this design,
dynamic power \emph{is} clock toggling. Clock gating attacks it directly, and the
synthesis-inserted ICG removes the most clock toggling ($-75.4\%$), which is
exactly why Opt2 has the lowest dynamic power overall.

\begin{table}[H]
\caption{Clock-network internal power, the gating target (mult workload, real).}
\label{tab:grp}\centering\footnotesize
\begin{tabular}{@{}lcc@{}}
\toprule
\textbf{Config} & \textbf{Clock-net power (\textmu W)} & \textbf{vs base} \\
\midrule
Baseline       & 49.63 & ref \\
Opt1 RTL       & 22.66 & $-54.3\%$ \\
Opt2 synth ICG & 12.19 & $-75.4\%$ \\
Opt3 both      & 28.75 & $-42.1\%$ \\
\bottomrule
\end{tabular}
\end{table}

\begin{figure}[t]
  \centering
  \includegraphics[width=\columnwidth]{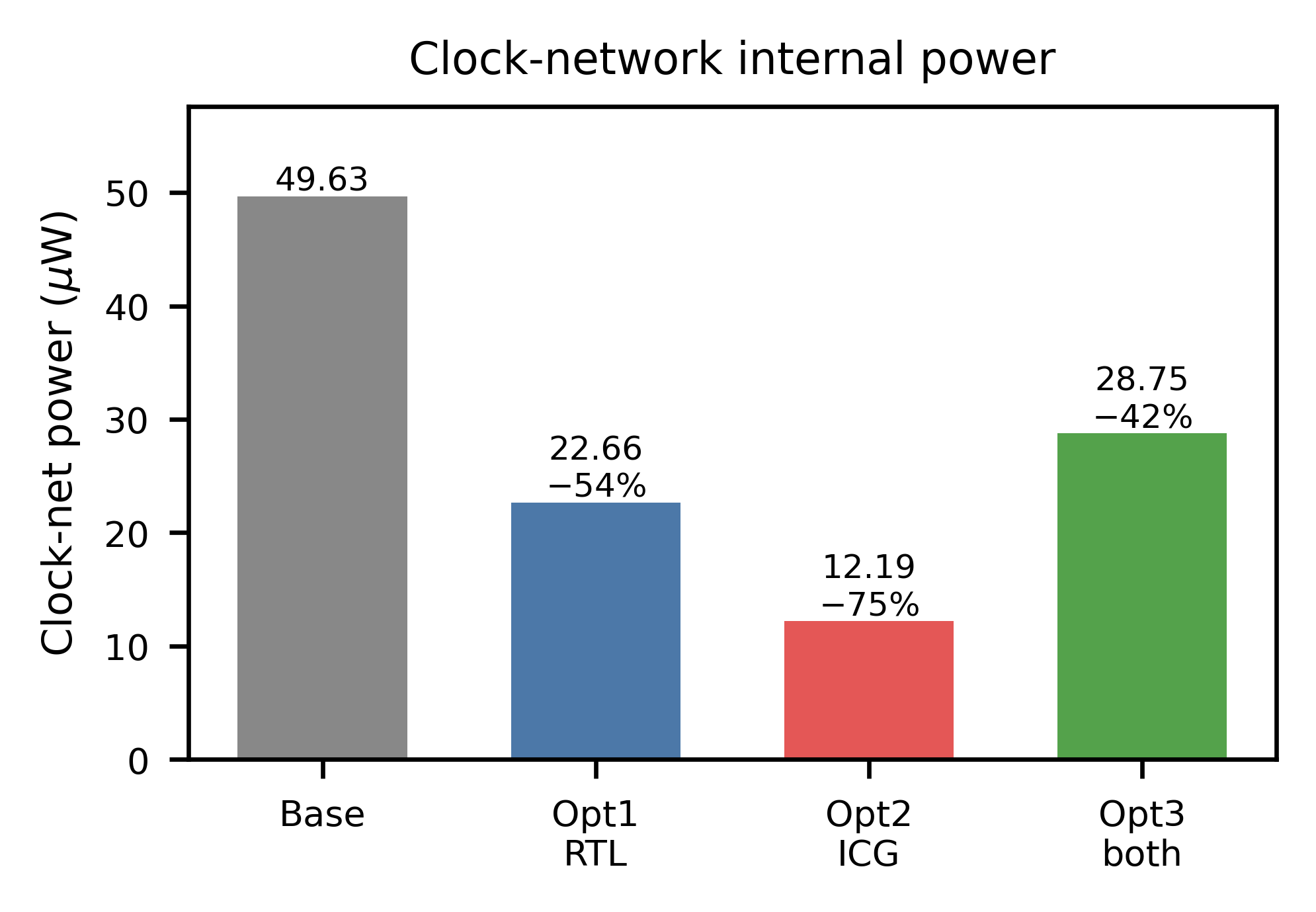}
  \caption{Clock-network internal power (real, mult workload): the gating target.
  ICG (Opt2) removes the most clock toggling ($-75.4\%$).}
  \label{fig:clocknet}
\end{figure}

\subsection{Corner sensitivity}\label{sec:corners}
\begin{figure*}[t]
  \centering
  \includegraphics[width=0.86\textwidth]{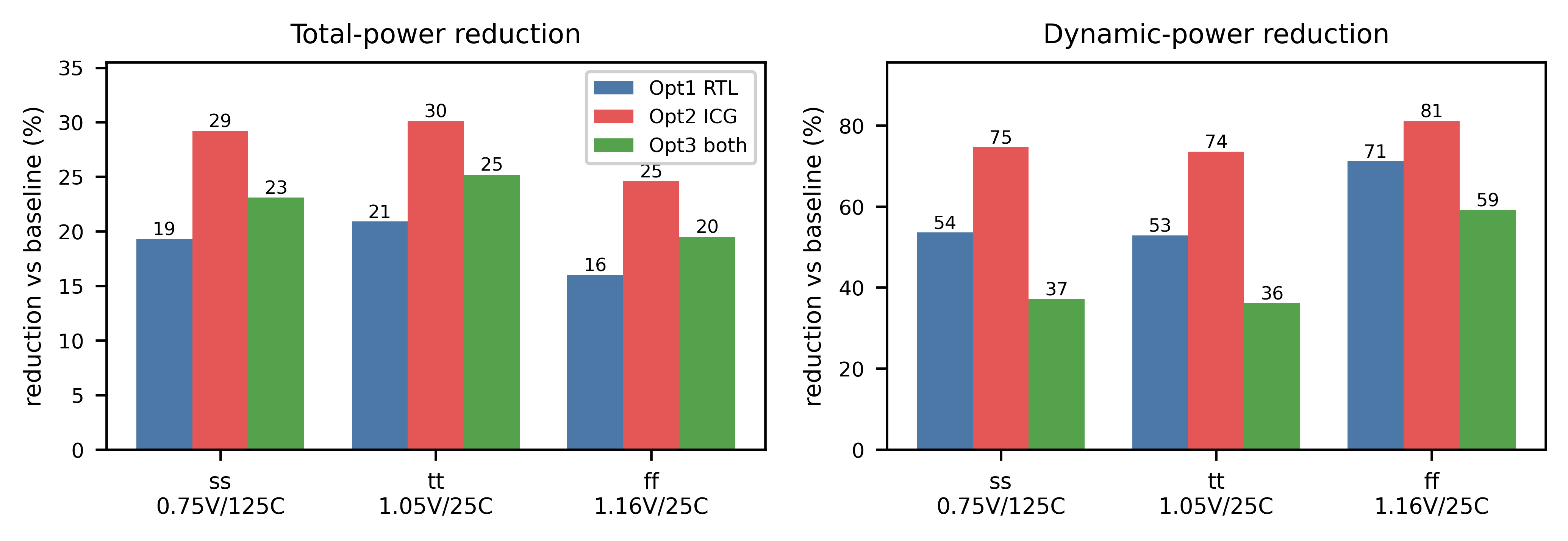}
  \caption{Corner sensitivity (real PrimeTime PX, mult workload): total- and
  dynamic-power reduction versus the ungated baseline at the ss/tt/ff corners.
  The gating benefit is corner-robust; ICG (Opt2) is best at every corner.}
  \label{fig:corners}
\end{figure*}

% [REAL] final/evidence/part6_multicorner_power.txt
Single-corner power can mislead, so we re-characterize average power at three
corners on the same netlists (activity is corner-independent; only the cell
power/leakage models change): worst-case \texttt{ss} (0.75\,V, 125\textcelsius),
typical \texttt{tt} (1.05\,V, 25\textcelsius), and fast \texttt{ff} (1.16\,V,
25\textcelsius). Fig.~\ref{fig:corners} reports the reduction versus the ungated
baseline. Two facts emerge. First, the gating benefit is \emph{corner-robust}:
ICG (Opt2) cuts total power by 25--30\% and dynamic power by 74--81\% at every
corner, and the ordering (Opt2 best) never changes. Second, in this 32\,nm
educational library leakage dominates total power at \emph{all} corners
(96--99\%), and rises steeply with supply voltage
($1.43\rightarrow16.2\rightarrow194.9$\,mW for the baseline at ss/tt/ff). The
total-power reduction therefore tracks a \emph{leakage} reduction (Opt2:
$-28\%/-30\%/-24\%$) that clock gating delivers through reduced area and simpler
register clouds (Table~\ref{tab:ppa}: $19474\rightarrow17745\,\mu m^2$), rather
than through the (large) dynamic saving, which is a small fraction of average
total power in this leakage-heavy regime. The dynamic saving instead dominates
\emph{active-mode energy}; the result argues for pairing clock gating with
leakage-reduction techniques (power gating, multi-$V_t$) when leakage dominates.

\subsection{Per-block breakdown}\label{sec:perblockbreak}
\begin{figure}[t]
  \centering
  \includegraphics[width=\columnwidth]{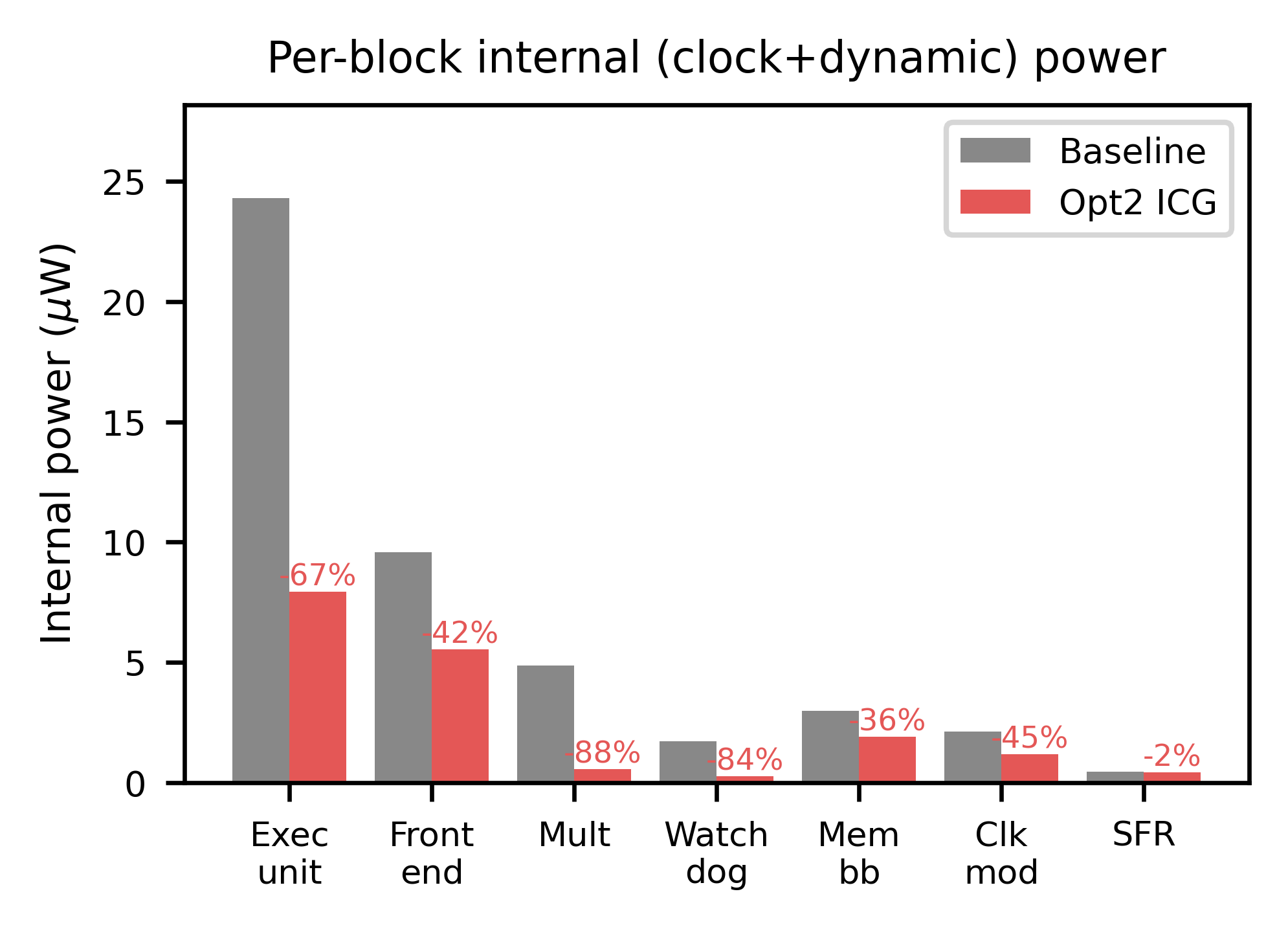}
  \caption{Per-block internal (clock$+$dynamic) power, baseline vs.\ ICG (Opt2),
  real PrimeTime hierarchy report. Gating cuts the often-idle multiplier and
  watchdog the most ($-88\%$, $-84\%$).}
  \label{fig:perblock}
\end{figure}

% [REAL] final/evidence/part8_perblock.txt (hierarchy-preserved synth + report_power -hierarchy)
To localize \emph{where} gating acts, we re-synthesized the design with hierarchy
preserved (\texttt{compile\_ultra -no\_autoungroup}) and used PrimeTime
\texttt{report\_power -hierarchy} to attribute internal (clock$+$dynamic) power to
each block (Fig.~\ref{fig:perblock}). The cut is concentrated exactly in the
blocks that are idle for much of normal operation: the hardware multiplier
($-88\%$) and the watchdog ($-84\%$), the blocks our proposal targets. The
always-active execution unit (register file $+$ ALU) and frontend still save
$67\%$ and $42\%$, because their registers are gated whenever they are not
updated. This per-block view confirms the optimization removes clock power where
expected rather than uniformly.

\subsection{Gate-level simulation caveat}\label{sec:caveat}
% [REAL] data from final/evidence/part2_netlist_analysis.txt
The behavioral RTL gate (Opt1/Opt3) is functionally correct in ideal RTL but its
\emph{netlist} does not simulate cleanly: the gated multiplier result is never
stored and \texttt{RESLO} reads \texttt{0x0000}. The CPU is \emph{not} frozen: we
confirmed via VCD that the datapath gate-enable toggles 21{,}926 times over the
run and the master clock toggles 100{,}000 times. The failure is confined to the
gated-register capture and is caused by a hold race: the hand-written latch+AND
produces a \emph{late} gated clock, so between the gated multiplier-result domain
and the ungated domain the result register samples the old/zero value.
Because the CPU runs normally and only the result-register capture is wrong, the
Opt1/Opt3 \emph{power} measurements (Tables~\ref{tab:ppa} and~\ref{tab:mwl})
remain valid: average power is determined by the pipeline and gating-enable
switching recorded in the VCD (21{,}926 enable toggles), not by the single
miscaptured result value.

This behavior is \emph{consistently reproducible for the discrete latch+AND
implementation under the evaluated synthesis, timing, and simulation flow}, not a
simulator-setting artifact: we tried eight simulation configurations
(zero/unit/full delay, \texttt{+nospecify}, \texttt{+notimingcheck},
several \texttt{+vcs+initreg} modes, and full SDF back-annotation), and all fail
identically. The discrete latch+AND gated clock is not protected by the
single-cell characterization and timing constraints that a tool-inserted ICG cell
carries, so the skew survives in this synthesis-level flow.
At the cell level, the Opt1 netlist contains 12 generic \texttt{LATCHX1} cells
(behavioral gate), whereas Opt2 uses 46 characterized \texttt{CGLPPRX2} integrated
clock-gating cells, which are designed to avoid this skew, so Opt2 passes 10/10
at gate level. This matches industry practice of using characterized ICG cells
rather than hand-written behavioral clock gates.

% [FIG 5] gate-level timing schematic (annotations are real data from part2_netlist_analysis.txt)
\begin{figure}[t]
  \centering
  \includegraphics[width=\columnwidth]{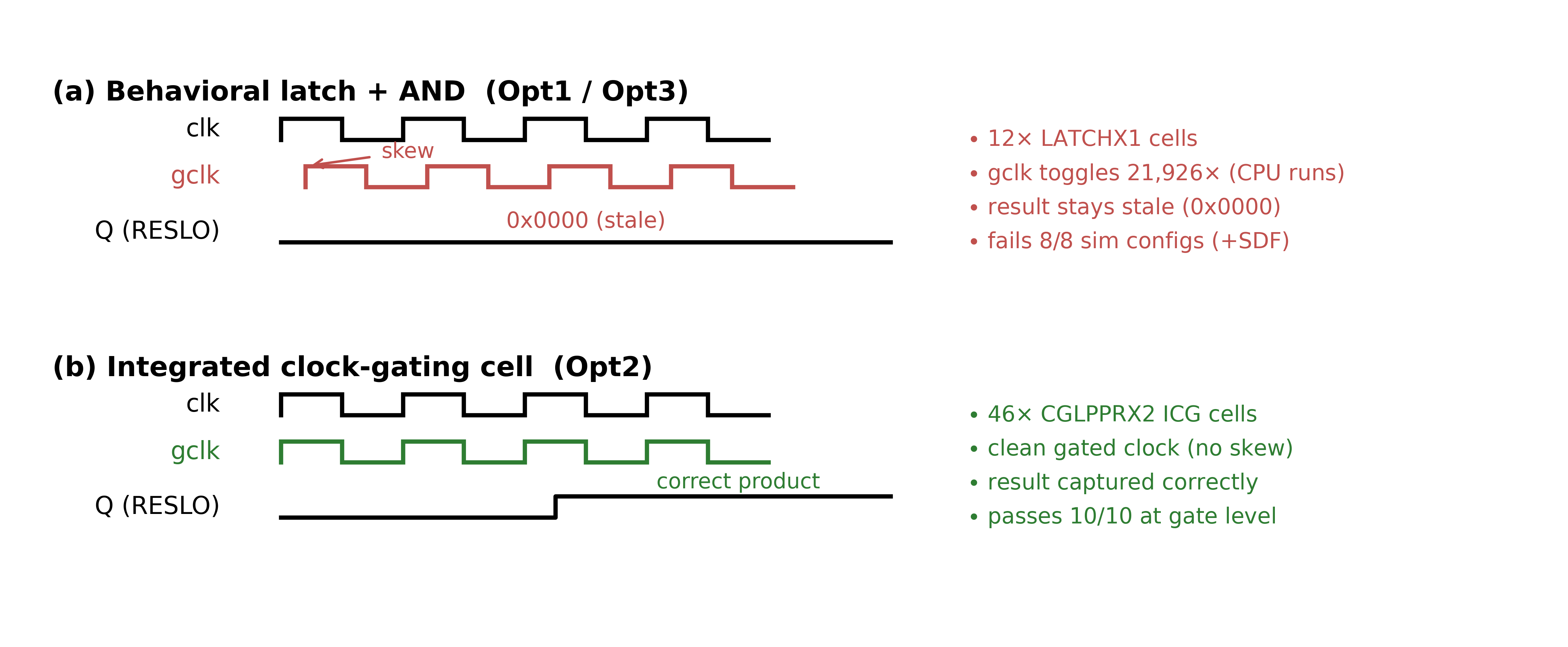}
  \caption{Root-cause schematic of the gate-level caveat (annotations are real
  measured data, \texttt{part2\_netlist\_analysis.txt}). (a) The hand-written
  latch+AND yields a \emph{late} gated clock; the result register races and keeps
  its stale value (\texttt{0x0000}) even though the CPU runs (21{,}926 gated-clock
  toggles). (b) The characterized ICG cell gates cleanly, so the result captures
  correctly. The same failure persists across eight simulation configurations
  including full SDF back-annotation.}
  \label{fig:wave}
\end{figure}

\begin{table}[H]
\caption{Cell evidence (from the synthesized netlists).}
\label{tab:cells}\centering\footnotesize
\begin{tabular}{@{}llc@{}}
\toprule
\textbf{Netlist} & \textbf{Gating cell} & \textbf{Count} \\
\midrule
Opt1 (behavioral) & \texttt{LATCHX1} (latch gate)   & 12 \\
Opt2 (synthesis)  & \texttt{CGLPPRX2} (ICG cell)    & 46 \\
\bottomrule
\end{tabular}
\end{table}

%==============================================================================
\section{Discussion}\label{sec:discuss}
%==============================================================================
Our results carry a practical message for low-power design on open-source cores.
Behavioral latch clock gates are attractive because they are visible and
controllable in RTL, and they are functionally correct, but they are not
gate-level-clean on a real library, which undermines sign-off via gate-level
simulation. Tool-inserted ICG cells achieve the larger power saving ($-29.2\%$
total, dynamic $49.6\rightarrow12.6$\,\textmu W) \emph{and} remain gate-level-clean.
The combined variant (Opt3) does not beat ICG alone, confirming that once ICG is
used the hand-written gate adds risk without benefit. The cost of gating, a small
$F_{max}$ reduction, is acceptable for the dominant power saving in
energy-constrained MCU/IoT use. The multi-workload measurements
(Section~\ref{sec:multiwl}) show the dynamic-power saving holds across diverse
activity profiles rather than for a single stimulus, and the clock-network
breakdown (Section~\ref{sec:perblock}) attributes it to suppressed clock
toggling, the intended mechanism. The corner study
(Section~\ref{sec:corners}) adds two cautions for practitioners. First, the
gating benefit is corner-robust, so a single-corner evaluation does not mislead
on the \emph{ranking} of strategies. Second, and less obviously, in a
leakage-dominated regime the headline total-power reduction is realized through
the area/leakage reduction that gating brings, not its dynamic saving; the
dynamic saving matters for active-mode energy. Where leakage dominates, clock
gating should therefore be paired with leakage-reduction techniques
(power gating, multi-$V_t$). Finally, the per-block breakdown
(Section~\ref{sec:perblockbreak}) shows the saving is not uniform: it concentrates
in the often-idle multiplier and watchdog. We also note that average power is
\emph{insensitive to instruction-level duty cycle} (a sweep of active/idle
instruction mix left it unchanged), because the pipeline stays clocked while
executing \texttt{NOP}s; meaningful gating savings come from whole-block idleness,
which is exactly what the per-block result captures.

%==============================================================================
\section{Conclusion}\label{sec:concl}
%==============================================================================
We presented a reproducible PPA study of three clock-gating strengths on the
open-source openMSP430 core with a 32\,nm library, and a gate-level
simulatability caveat: hand-written behavioral latch clock gates are RTL-correct
but not gate-level-simulatable (the gated result reads zero, despite 21{,}926
enable toggles), whereas tool-inserted ICG cells are clean and deliver the largest
power saving. The saving is corner-robust across ss/tt/ff (ICG: total $-25$ to
$-30\%$, dynamic $-74$ to $-81\%$); in this leakage-dominated regime the
total-power win is realized through gating's area/leakage reduction, while the
dynamic saving dominates active-mode energy. The concrete recommendation is to
prefer tool-inserted ICG over behavioral gates for gate-level-clean low-power
design, paired with leakage reduction where leakage dominates.

%==============================================================================
\section*{Artifact Availability}\label{sec:artifact}
%==============================================================================
\ifreview
The complete reproducible flow (RTL clock-gating switch, Design Compiler
synthesis scripts, PrimeTime PX power scripts, and the self-checking testbench)
will be released as an open-source artifact upon acceptance. % link omitted for review (anonymity-safe)
\else
The RTL switch, full Design Compiler synthesis flow, PrimeTime PX power scripts,
and self-checking testbench are available at
\url{https://github.com/yanyana117/openmsp430-low-power-study-full}.
\fi

%------------------------------------------------------------------- BIB
\bibliographystyle{IEEEtran}
\bibliography{references}

\end{document}